\documentclass[conference]{IEEEtran}
\IEEEoverridecommandlockouts
\usepackage{cite}
\usepackage{amsmath,amssymb,amsfonts}
\usepackage{graphicx}
\usepackage{textcomp}
\usepackage{xcolor}
\usepackage{bm}
\usepackage{float}
\usepackage{url}
\def\BibTeX{{\rm B\kern-.05em{\sc i\kern-.025em b}\kern-.08em44
    T\kern-.1667em\lower.7ex\hbox{E}\kern-.125emX}}
    
\usepackage{algorithm}
\usepackage[noend]{algpseudocode}
\makeatletter
\def\BState{\State\hskip-\ALG@thistlm}
\makeatother

\usepackage{color}
\usepackage{tikzscale}

\usepackage{multicol}

\DeclareGraphicsExtensions{.eps}
\usepackage{lipsum}

\usepackage{pgfplots}

\usepackage{tikz}
\usetikzlibrary{shapes.geometric}

\usetikzlibrary{fit,shapes.geometric}
\usetikzlibrary{shapes,arrows,chains}
\pgfdeclarelayer{signal}
\pgfsetlayers{signal,main}
\tikzset{pics/.cd,
  SBS/.style={code={
      \begin{scope}[local bounding box=#1]
      \fill [pic actions/.try] (-1,0) -- (-1/2,3) -- (1/2, 3) -- (1,0) -- cycle;
      \fill [pic actions/.try] (-1/16,2) rectangle (1/16,4);
      \fill [pic actions/.try] (0,4) circle [radius=1/4];
      \foreach \i in {-1,1}
        \fill [shift=(90:4), xscale=\i]
          \foreach \r in {1,3/2,2}{
            (-45:\r) arc (-45:45:\r) -- (45:\r-1/10)
            arc(45:-45:\r-1/10) -- cycle
          };
       \end{scope}
  }},
  SU/.style={code={
      \begin{scope}[local bounding box=#1]
      \fill [pic actions/.try] (-5/2,3/2) -- (-5/2,3) -- (1/2, 3) -- (1/2,3/2) -- cycle;
      \fill [pic actions/.try] (-1/16,2) rectangle (1/16,4);
      \fill [pic actions/.try] (0,4) circle [radius=1/4];
      \fill [red,pic actions/.try] (-2,2) circle [radius=1/5];
      \fill [green,pic actions/.try] (-2,5/2) circle [radius=1/5];
      \fill [white,pic actions/.try] (-0.75,2.25) circle [radius=1/3];
      \foreach \i in {-1,1}
        \fill [shift=(90:4), xscale=\i]
          \foreach \r in {0.75,1}{
            (-45:\r) arc (-45:45:\r) -- (45:\r-1/10)
            arc(45:-45:\r-1/10) -- cycle
          };
       \end{scope}
  }},
  SIGNAL/.style={code={
    \begin{scope}[local bounding box=#1]
      \fill [pic actions/.try]
      (0,-3) -- (-1,1/2) -- (1/8,1/4) -- (0,3) -- (1,-1/2) -- (-1/8,-1/4)
      -- cycle;
    \end{scope}
  }}
}
\colorlet{sky blue}{blue!60!cyan!75!black}
\colorlet{dark blue}{blue!50!cyan}
\colorlet{chameleon}{olive!75!green}
\tikzset{signal/.style={draw=gray, line width=0.2em, dashed}}

\begin{document}

\title{Stochastic Geometry Analysis and Design of Wireless Powered MTC Networks\\
\thanks{The work presented in this paper was carried out within the framework of the project 5G\&B RUNNER-UPC (TEC2016-77148-C2-1-R (AEI/FEDER, UE)), the research network RED2018-102668-T Red COMONSENS and the~FPI grant BES-2017-079994, funded by the Spanish Ministry of Science, Innovation and Universities; and~the grant 2017 SGR 578, funded by the Catalan Government (AGAUR, Secretaria d'Universitats i Recerca, Departament d'Empresa i Coneixement, Generalitat de Catalunya).}}

\author{\IEEEauthorblockN{Sergi Liesegang, Olga Mu\~noz-Medina, and Antonio Pascual-Iserte}
\IEEEauthorblockA{Dept. Signal Theory and Communications - Universitat Polit\`ecnica de Catalunya (UPC), Barcelona, Spain}
Emails: \{sergi.liesegang, olga.munoz, antonio.pascual\}@upc.edu}

\maketitle

\begin{abstract}
Machine-type-communications (MTC) are being crucial in the development of next generation mobile networks. Given that MTC devices are usually battery constrained, wireless power transfer (WPT) and energy harvesting (EH) have emerged as feasible options to enlarge the lifetime of the devices, leading to wireless powered networks. In that sense, we consider a setup where groups of sensors are served by a base station (BS), which is responsible for the WPT. Additionally, EH is used to collect energy from the wireless signals transmitted by other sensors. To characterize the energy obtained from both procedures, we model the sporadic activity of sensors as Bernoulli random variables and their positions with repulsive Mat\'ern cluster processes. This way, the random activity and spatial distribution of sensors are introduced in the analysis of the energy statistics. This analysis can be useful for system design aspects such as energy allocation schemes or optimization of idle-active periods, among others. As an example of use of the developed analysis, we include the design of a WPT scheme under a proportional fair policy. 
\end{abstract}
\section{Introduction}
\label{sec:1}

Machine-type-communications (MTC) have a prominent role in the evolution of future mobile systems \cite{Che17}. They define a type of networks where sets of devices communicate with low human supervision and where the number of connected terminals is expected to grow exponentially \cite{Wan17}.  
In many MTC applications, as those under the umbrella of the Internet-of-Things \cite{Raz17}, the devices can be energy constrained, specially if the charging or replacement of batteries is difficult \cite{Boc16}.  

That is why strategies such as wireless power transfer (WPT) and energy harvesting (EH) are interesting candidates to improve the lifetime of devices and enable wireless powered MTC networks \cite{Bi15,Cho17,Pon18}. WPT represents the transfer of energy through dedicated signals, usually in the downlink (DL) by a base station (BS), and EH consists in collecting energy from the signals transmitted by other devices in the same region.

In this work, we consider a setup with a set of sensors and a serving BS equipped with multiple antennas. The BS is in charge of the WPT in the DL, i.e. it transfers energy to the sensors. Also, given the high spatial density of these devices in networks such as \textit{massive} MTC \cite{Boc16}, each of them will take advantage of this by harvesting energy from signals coming from other sensors when they transmit (TX) in the uplink (UL). 

Given that in MTC networks the positions of the sensors are considered random and normally unknown, in this paper we make use of stochastic geometry to model their spatial distribution \cite{ElS13}. In particular, we consider that they are represented by repulsive Mat\'ern cluster processes (MCPs). Besides, for a more realistic analysis, the intermittent activities of the sensing devices are also included through Bernoulli random variables (RVs). This allows us to derive and statistically characterize both harvested and transferred energies. This is indeed the main novelty of this work. Finally, in order to optimize the collected energy, in this paper we design an energy allocation scheme with proportional fairness through the WPT. This way, sensors receive a similar amount of energy over time. 

The remainder of this paper is structured as follows. Section II describes the system model. In Section III, the WPT and the EH energies are statistically characterized. Section IV is devoted to the WPT design and numerical simulations are shown in Section V. Conclusions are presented in Section VI.

\section{System Model}
\label{sec:2}

Throughout this work, we consider a scenario with a set of sensors randomly located in space and served by a BS. In particular, we consider that these devices are grouped together (organized) in $K$ clusters, each one spatially represented by a disk of radius $R_k$ centered at $\bm{c}_k \in \mathbb{R}^2$, with $k \in \{1,\ldots,K\}$. Both $R_k$ and $\bm{c}_k$ are assumed to be known, but not the sensors positions. Each cluster has a density of sensors $\lambda_k$ and sensors in the same cluster are at minimum distance $d_{\textrm{min}}$ between each other. We also assume that no devices are located outside these delimited regions. Note that clustering is essential to structure MTC networks and improve their performance, e.g. scalability, coverage, throughput and energy consumption \cite{Leu15}.

The positions of the sensors in cluster $k$ can be represented by a repulsive MCP $\Upsilon_k$ of intensity $\lambda_k$ and minimum distance $d_{\textrm{min}}$ \cite{Chu15}. Each of these processes is defined from a general MCP $\Psi_k$ of intensity $\delta_k$ and without minimum distance, where devices are uniformly distributed in the circle of radius $R_k$ around the center $\bm{c}_k$. Then, in order to ensure the minimum sensor distance, a dependent thinning is applied to each MCP $\Psi_k$ \cite{And16}. This yields the repulsive process $\Upsilon_k$ with density of devices \cite{Chi13}:%
\begin{equation}
\lambda_k = \left(1 - e^{-\delta_k \pi d_{\textrm{min}}^2}\right)/\left(\pi d_{\textrm{min}}^2\right). %
\label{eq:1}
\end{equation}

For $d_{\textrm{min}} \to 0$, $\Upsilon_k$ can be safely approximated by a MCP $\Phi_k$ with the same boundaries (i.e. disk of radius $R_k$ and center $\bm{c}_k$) and density of sensors $\lambda_k$ \cite{ElS16}. Note that the number of devices in each $\Phi_k$ is a Poisson RV with parameter $\lambda_k \pi R_k^2$, i.e. $\vert \Phi_k \vert \sim $ \textit{Poisson}($\lambda_k \pi R_k^2$), and that the processes $\Phi_k$ modeling different clusters are independent between each other.

\begin{figure}[t] 
\centerline{\includegraphics[height=4.5cm,keepaspectratio]{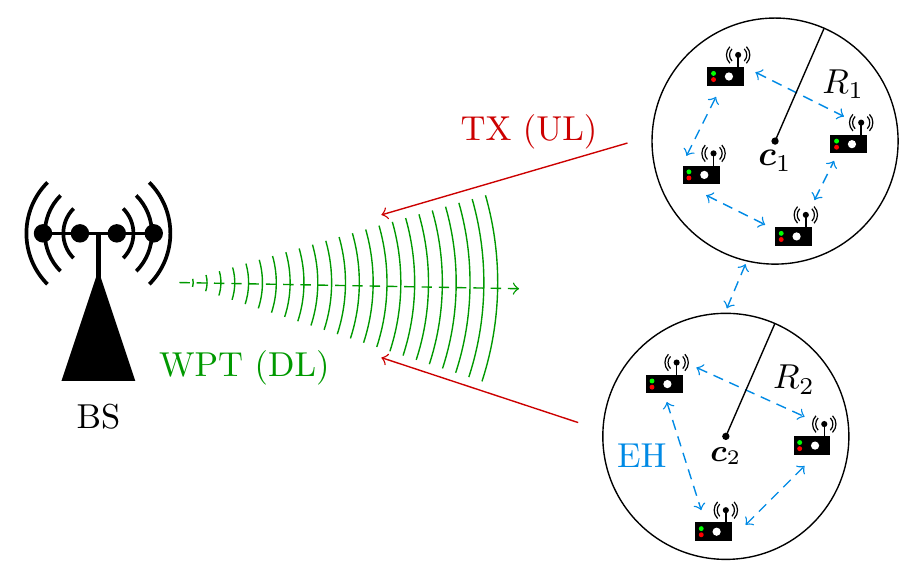}} 
\vspace{-2mm}
\caption{Illustrative scenario with $K = 2$ and $M = 4$.}
\label{fig:1}
\vspace{-4mm}
\end{figure}

We also consider that the BS is equipped with $M$ antennas. This will result in an array gain that can counteract the high propagation losses and better direct the energy to where it is needed when the BS applies WPT. In addition to that, we assume that sensors are single-antenna devices due to required simplicity and low cost. An illustrative example of a scenario with $K = 2$ and $M = 4$ is depicted in Fig. \ref{fig:1}.

In the following, given the limited capability of the sensors, we assume that WPT and EH are not performed simultaneously, yet in a half-duplex way. Thereby, considering that time is divided into frames of duration $T_f$, WPT will occupy the first $T_d$ seconds and EH the other $T_u$ such that $T_f = T_d + T_u$.

\begin{figure}[t]
\centerline{\includegraphics[width=0.43\textwidth]{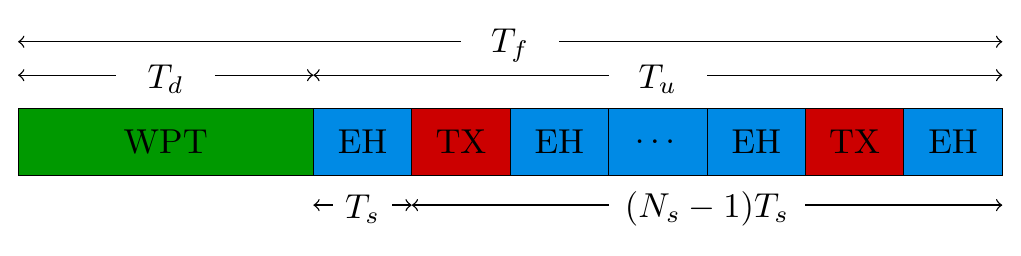}}
\vspace{-4mm}
\caption{Example of the frame structure.}
\label{fig:2}
\vspace{-4mm}
\end{figure}

Following the previous reasoning, whenever a sensor is transmitting, it will be unable to harvest energy from the signals transmitted by other active sensors, i.e. TX and EH are not simultaneous. Then, the period $T_u$ is also divided into $N_s$ slots of duration $T_s$ such that $T_u = N_s T_s$. Hence, some slots will be dedicated to TX and the rest to EH. In addition, given that sensors actually transmit in a sporadic way \cite{3GPP45820}, we consider that the probability of being active at any time slot is $p_{\textrm{act}}$. An example of the frame structure is shown in Fig. \ref{fig:2}.

As a result, the average energy that the sensors in cluster $k$ will receive from WPT and EH at the end of each frame is %
\begin{equation}
E_k = E_{k,\textrm{WPT}} + E_{k,\textrm{EH}},%
\label{eq:2}
\end{equation}
where $E_{k,\textrm{WPT}}$ refers to the average energy obtained from WPT and $E_{k,\textrm{EH}}$ is the average energy collected from EH thanks to the other active sensors that are transmitting. In the upcoming section, these energies will be properly characterized.

\section{Energy Characterization}
\label{sec:3}

In order to statistically characterize $E_{k,\textrm{WPT}}$ and $E_{k,\textrm{EH}}$, we first model the received signals at the sensors side. Later, with stochastic geometry tools, we derive the expressions for both average energies. This represents the core of our paper. Based on that, in Section \ref{sec:4}, we design an energy allocation scheme for WPT where the average energies $E_k$ are maximized under a BS power constraint and a proportional fair policy \cite{Vis02}.


\subsection{Energy received from WPT}
\label{sec:3.1}

The signal received from the BS by a sensor randomly chosen in the set of cluster $k$ and located at $\hat{\bm{x}}_k \in \Phi_k$ is %
\begin{equation}
\hat{y}_k^{\textrm{WPT}} = \hat{\bm{h}}_k^{\textrm{H}} \bm{s} + \hat{n}_k \in \mathbb{C},%
\label{eq:3}
\end{equation}
where $\hat{\bm{h}}_k \in \mathbb{C}^M$ is the channel of the sensor with respect to (w.r.t.) the BS, $\bm{s}\in \mathbb{C}^M$ is the BS transmit signal with zero-mean and covariance $\bm{Q} =  \textrm{E}[\bm{s}\bm{s}^{\textrm{H}}] \in \mathbb{C}^{M \times M}$, and $\hat{n}_k \in \mathbb{C}$ is the corresponding noise with zero-mean and power $\sigma_n^2$. Ignoring the negligible energy coming from this last term \cite{Zho13}, the received energy can be expressed as 
\begin{equation}
E_{k,\textrm{WPT}} = T_d \textrm{E}[\hat{\bm{h}}_k^{\textrm{H}} \bm{Q} \hat{\bm{h}}_k] = T_d \textrm{tr}(\bm{Q}\textrm{E}[\hat{\bm{h}}_k \hat{\bm{h}}_k^{\textrm{H}}]) = T_d \textrm{tr}(\bm{Q}\bm{C}_k),%
\label{eq:4}
\end{equation}
where the entries of $\bm{C}_k = \textrm{E}[\hat{\bm{h}}_k \hat{\bm{h}}_k^{\textrm{H}}] \in \mathbb{C}^{M \times M}$ are given by %
\begin{equation}
[\bm{C}_k]_{i,j} = \textrm{E}[\hat{h}_{k,i} \hat{h}_{k,j}^*], \quad 1 \leq i \leq M, \quad 1 \leq j \leq M,%
\label{eq:5}
\end{equation}
which average the sensor selection over the process $\Phi_k$. 

Considering a power law path-loss, the sensor channel in \eqref{eq:3} reads as
\begin{equation}
\hat{\bm{h}}_k = \hat{d}_k^{-\alpha/2} \hat{g}_k \bm{v}(\hat{\vartheta}_k),
\label{eq:6}
\end{equation}
where $\hat{d}_k$ is the distance of the sensor located at $\hat{\bm{x}}_k \in \Phi_k$ to the BS, $\alpha$ is the decay exponent, $\hat{g}_k$ is the fading coefficient with zero-mean and variance $\sigma_g^2$, $\bm{v}(\cdot) = [v_1,\ldots,v_M] \in \mathbb{C}^M$ is the steering vector and $\hat{\vartheta}_k$ is the steering direction \cite{Van04}. 

\begin{figure}[t] 
\centerline{\includegraphics[width=0.35\textwidth]{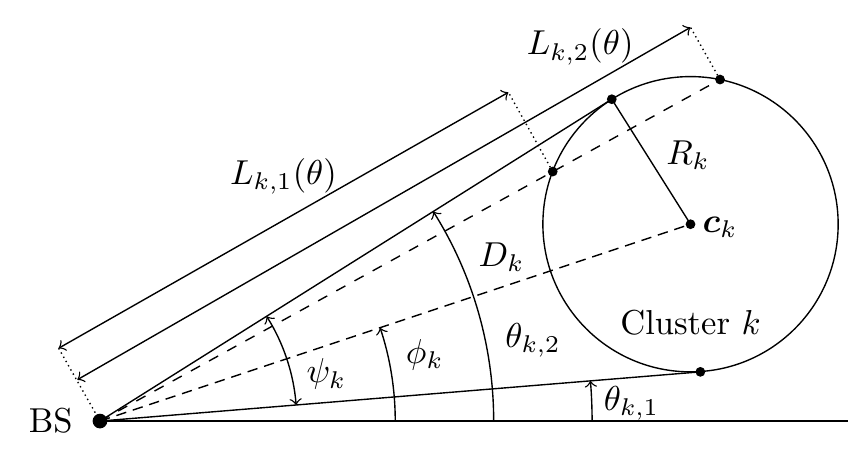}}
\vspace{-4mm}
\caption{Geometry between the BS and the $k$-th cluster.}
\label{fig:3}
\vspace{-4mm}
\end{figure}

As a result, the entries of $\bm{C}_k$ defined in \eqref{eq:5} yield
\begin{equation}
\begin{split}
[\bm{C}_k]_{i,j} &= \textrm{E}[\hat{d}_k^{-\alpha} \hat{g}_k^2v_i(\hat{\vartheta}_k)v_j(\hat{\vartheta}_k)^*] \\
&\stackrel{\textrm{(a)}}{=} \sigma_g^2 \textrm{E}[\hat{d}_k^{-\alpha}v_i(\hat{\vartheta}_k)v_j(\hat{\vartheta}_k)^*] \\
&\stackrel{\textrm{(b)}}{=}  \frac{\sigma_g^2}{\pi R_k^2} \int_{\theta_{k,1}}^{\theta_{k,2}}  \left( \int_{L_{k,1}(\theta)}^{L_{k,2}(\theta)} \frac{r}{r^{\alpha}} dr \right) v_i(\theta)v_j(\theta)^* d\theta \\
&= \frac{\sigma_g^2}{\pi R_k^2} \int_{\theta_{k,1}}^{\theta_{k,2}}  I_k(\theta) v_i(\theta)v_j(\theta)^* d\theta,
\end{split}
\label{eq:7}
\end{equation}
where (a) follows from the independence of fading and spatial process, and (b) from the uniform spatial distribution of the sensors and the change to polar coordinates. As depicted in Fig. \ref{fig:3}, $\theta_{k,1}$ and $\theta_{k,2}$ are the two angles corresponding to the tangent lines from the BS to cluster $k$, i.e.
\begin{equation}
\theta_{k,1} = \phi_{k} - \psi_k/2, \quad \theta_{k,2} = \phi_k + \psi_k/2,
\label{eq:8}
\end{equation}
where $\psi_k = 2 \arcsin(R_k/D_k)$ is the angle between these two tangent lines, $D_k$ is the distance between the cluster center $\bm{c}_k$ and the BS, and $\phi_k$ is the associated angle, as shown in Fig. \ref{fig:3}. Note that the integral term $I_k(\theta)$ in \eqref{eq:7} depends on the decay exponent $\alpha$ and can be expressed as
\begin{equation}
I_k(\theta) = \left\{
        \begin{array}{ll}
            \ln \left(L_{k,2}(\theta)/L_{k,1}(\theta)\right), & \alpha = 2, \\
            \frac{1}{2 - \alpha} (L_{k,2}(\theta)^{2 - \alpha} - L_{k,1}(\theta)^{2 - \alpha}), & \alpha > 2,
        \end{array}
    \right.
\label{eq:9}
\end{equation}
where $L_{k,1}(\theta) \leq L_{k,2}(\theta)$ are the distances to the BS of the intersections between the lines determined by $\theta_{k,1}$ and $\theta_{k,2}$, and the boundaries of the $k$-th cluster. They are computed as the roots of the polynomial:
\begin{equation}
p(\theta) = r^2 - 2D_k r\cos(\theta - \phi_k) + D_k^2 - R_k^2.
\label{eq:10}
\end{equation}

Finally, as there is no closed-form expression for the integral in \eqref{eq:7}, it is computed numerically. This way, we can obtain the values for the matrix $\bm{C}_k$ formulated in \eqref{eq:4} and \eqref{eq:5}.

\subsection{Energy received from EH}
\label{sec:3.2}

Regarding the received signal from the other sensors, we also need to take into account their spatial distribution. To ease of notation, we again consider that the sensor under study is located at $\hat{\bm{x}}_{k}$ inside cluster $k$. For a power law path-loss, the received signal at a given slot $n \in \{1,\ldots,N_s\}$ is given by %
\begin{equation}
\hat{y}_k^{\textrm{EH}} [n] = \sum_{k' = 1}^K \hat{y}_{k,k'}^{\textrm{EH}}[n] (1 - \beta_{\hat{\bm{x}}_{k}} [n])  +  \hat{w}_k \in \mathbb{C}, %
\label{eq:11} 
\end{equation}
with %
\begin{equation}
\hat{y}_{k,k'}^{\textrm{EH}} [n] = \sum_{\bm{x} \in \Phi_{k'} \backslash \{\hat{\bm{x}}_{k}\} } \hat{\rho}_{k,\bm{x}}^{-\alpha/2} \hat{\gamma}_{k,\bm{x}} [n] \tau_{\bm{x}} [n] \beta_{\bm{x}} [n] \in \mathbb{C}, %
\label{eq:12}
\end{equation} 
where $\hat{\rho}_{k,\bm{x}}$ is the distance between $\bm{x}$ and $\hat{\bm{x}}_{k}$, $\hat{\gamma}_{k,\bm{x}} [n] \in \mathbb{C}$ is the fading coefficient with zero-mean and variance $\sigma_{\gamma}^2$, $\tau_{\bm{x}} [n] \in \mathbb{C}$ is the sensor transmit signal with zero-mean and power $P_{\tau}$, and $\hat{w}_k [n]$ is the noise with zero-mean and power $\sigma_w^2$. $\beta_{\bm{x}} [n] \sim $ \textit{Ber}($p_{\textrm{act}}$) are independent Bernoulli RVs introduced to account for the random activity of the sensor located at $\bm{x}$, i.e. $p_{\textrm{act}}$ is the probability that a sensor is active and transmitting during the time slot $n$ \cite{Lie18}. In that sense, $1 - \beta_{\hat{\bm{x}}_{k}} [n]$, with $\beta_{\hat{\bm{x}}_{k}} [n] \sim$ \textit{Ber}($p_{\textrm{act}}$) independent of the rest, represents the fact that the sensor located at $\hat{\bm{x}}_{k}$ can only harvest energy from the other active devices when it is not transmitting. 

As a result, neglecting again the noise energy, the average energy harvested by a sensor at the $k$-th cluster reads as \cite{Chi13} 
\begin{align}
E_{k,\textrm{EH}} &= \textrm{E}\left[\sum_{n = 1}^{N_s}  T_s \left(\hat{y}_{k}^{\textrm{EH}} [n]\right)^2  \right]\\
&= N_s T_s P_{\tau} \sigma_{\gamma}^2 p_{\textrm{act}} (1 - p_{\textrm{act}}) \underbrace{\textrm{E}\left[\sum_{k' = 1}^K \sum_{\bm{x} \in \Phi_{k'} \backslash \{\hat{\bm{x}}_{k}\} } \hat{\rho}_{k,\bm{x}}^{-\alpha}\right]}_{\triangleq \eta_k}, \nonumber  
\label{eq:13}
\end{align}
which follows from the independence of fading, transmit signal and activity RVs. Note that $\eta_k$ can be decomposed as %

\begin{equation*}
\eta_k = \textrm{E}\left[\sum_{k' = 1}^K \sum_{\bm{x} \in \Phi_{k'} \backslash \{\hat{\bm{x}}_{k}\}} \hat{\rho}_{k,\bm{x}}^{-\alpha}\right] = \sum_{k' = 1}^K\textrm{E}\left[ \sum_{\bm{x} \in \Phi_{k'} \backslash \{\hat{\bm{x}}_{k}\}} \hat{\rho}_{k,\bm{x}}^{-\alpha}\right]
\end{equation*}
\begin{equation}
\hspace{-10mm} = \underbrace{\textrm{E}\left[ \sum_{\bm{x} \in \Phi_{k} \backslash \{\hat{\bm{x}}_{k}\}} \hat{\rho}_{k,\bm{x}}^{-\alpha}\right]}_{\triangleq \eta_{k,\textrm{intra}}} + \underbrace{\sum_{k' \neq k}\textrm{E}\left[ \sum_{\bm{x} \in \Phi_{k'}} \hat{\rho}_{k,\bm{x}}^{-\alpha}\right]}_{\triangleq \eta_{k,\textrm{inter}}},
\label{eq:14}
\end{equation}
where $\eta_{k,\textrm{intra}}$ is the sum of path-loss within the same cluster (i.e. intra-cluster) and $\eta_{k,\textrm{inter}}$ is the sum of path-loss from the rest (i.e. inter-cluster). The former is expressed in an analytic closed-form, yet an approximation is used for the latter.

With the help of Campbell's theorem \cite{And16}, the sum of intra-cluster path-loss (without the harvesting sensor) can be written as
\begin{equation}
\eta_{k,\textrm{intra}} = (\lambda_k \pi R_k^2 - 1) \int_{d_{\textrm{min}}}^{2 R_k} \rho^{-\alpha} f_{k} (\rho) d \rho,
\label{eq:15}
\end{equation}
where $f_{k}(\rho)$ is the distance distribution in cluster $k$ \cite{Chi13}:
\begin{equation}
f_{k}(\rho) = \frac{4 \rho}{A_k \pi R_k^2} \left( \arccos \left(\frac{\rho}{2 R_k}\right) - \frac{\rho}{2 R_k} \sqrt{1 - \frac{\rho^2}{4 R_k^2}}\right),
\label{eq:16}
\end{equation}
in the interval $d_{\textrm{min}}\leq \rho_k \leq 2R_k$ and $0$ otherwise. $A_k$ is the normalization factor that ensures the distribution has unit area.

On the other hand, the sum of inter-cluster path-loss can be difficult to model as the position of the sensor under study is also random, i.e. $\hat{\bm{x}}_{k} \in \Phi_k$. Instead of focusing on the energy harvested by a sensor located in cluster $k$, we concentrate on the energy received at the cluster center, i.e. $\hat{\bm{x}}_{k} = \bm{c}_{k}$. 

This way, following the discussion in Section \ref{sec:3.1}, using Campbell's theorem, $\eta_{k,\textrm{inter}}$ can be approximated as \cite{ElS16}:
\begin{equation}
\eta_{k,\textrm{inter}} \approx \tilde{\eta}_{k,\textrm{inter}} = \sum_{k' \neq k} \lambda_{k'} \int_{\theta_{k,1}}^{\theta_{k,2}}  I_k(\theta) d\theta,
\label{eq:17}
\end{equation}
where $\theta_{k,1}$, $\theta_{k,2}$ and $I_k(\theta)$ are those defined in \eqref{eq:8} and \eqref{eq:9}, but observed from the point of view of the cluster center $\bm{c}_k$. 

The use of the approximation in \eqref{eq:17} can be justified by the small channel gains and the high attenuation in the sensor-to-sensor link. 
In that sense, clusters far apart can be seen approximately as a point and, thus, the energy that a sensor harvests from other clusters will be similar over its own cluster. This is verified through simulations in Section V, where we compare the actual value $\eta_{k,\textrm{inter}}$ and the approximation $\tilde{\eta}_{k,\textrm{inter}}$. 

\section{Energy Allocation}
\label{sec:4}

The previous analysis can be useful in the design of MTC systems. As an example, in this section we present the design of an energy allocation scheme for WPT from the BS under a proportional fair policy that can lead to a wireless powered network. To that aim, we maximize the sum of the logarithm of the sensors long-term collected energy under a BS total power constraint \cite{Vis02}. Note that this is equivalent to maximizing the product of the sensors received energy, which results in a fair distribution of energy within the system \cite{Tse05}.

We start by considering an observation time of $T$ frames. Thus, now the received energy $E_k$ in \eqref{eq:2} depends on the frame $t \in \{1,\ldots,T\}$, i.e. $E_k (t)$ is the received energy at the $k$-th cluster after the $t$-th frame. Following the previous analysis: 
\begin{equation}
E_k(t) = T_d \textrm{tr}(\bm{Q}(t)\bm{C}_k) + T_u P_{\tau} \sigma_{\gamma}^2 p_{\textrm{act}} (1 - p_{\textrm{act}}) \eta_k,
\label{eq:18}
\end{equation}
where the covariance matrix $\bm{Q}(t)$ is allowed to change over the different frames to optimize the accumulated energy. In fact, note that this matrix determines the energy coming from the BS that arrives at the different sensors of each cluster and, therefore, it will be the design variable of the system. 



As a result, the optimization problem can be defined as \cite{Tek13}
\vspace{-2mm}
\begin{equation}
\bm{Q}^{\star} (t) = \underset{\bm{Q}(t)\succeq \bm{0}}{\textrm{argmax}} \, \, \sum_{k = 1}^K \log T_k(t) \quad \textrm{s.t.}\ \textrm{tr}(\bm{Q}(t)) \leq P_{\textrm{tx}},
\label{eq:19}
\end{equation}
where $\bm{Q}(t) \succeq \bm{0}$ expresses that the matrix $\bm{Q}(t)$ must be positive semi-definite by definition, and $P_{\textrm{tx}}$ is the total transmit power available at the BS. The terms $T_k(t)$ are the exponentially averaged received energies \cite{Rub14}, i.e. 
\begin{equation}
T_k(t) = \left(1 - \frac{1}{T_c}\right) T_k (t - 1) + \frac{1}{T_c} E_k (t),
\label{eq:20}
\end{equation}
with $T_c$ being the effective length of the exponential impulse response of the averaging filter in terms of number of frames.

As shown in the Appendix, for a sufficiently large window duration $T_c$, the problem in \eqref{eq:19} is equivalent to \cite{Ngu06}: 
\begin{equation}
\bm{Q}^{\star}(t) = \underset{\bm{Q}(t) \succeq \bm{0}}{\textrm{argmax}} \, \, \sum_{k = 1}^K w_k(t) E_k(t) \quad \textrm{s.t.}\ \textrm{tr}(\bm{Q}(t)) \leq P_{\textrm{tx}},
\label{eq:21}
\end{equation}
where the weights $w_k(t) = 1/T_k (t - 1)$ scale the individual $E_k(t)$ such that a higher priority is given to the sensors that have accumulated less energy during past frames. 

The solution to the previous problem is given by \cite{Tse05}
\begin{equation}
\bm{Q}^{\star}(t) = P_{\textrm{tx}}\bm{v}_{\textrm{max}}(t)\bm{v}_{\textrm{max}}^{\textrm{H}}(t),
\label{eq:22}
\end{equation}
where $\bm{v}_{\textrm{max}}(t)$ is the eigenvector of the largest eigenvalue from
\begin{equation}
\bm{C}(t) = T_d \sum_{k = 1}^K w_k(t) \bm{C}_k \in \mathbb{C}^{M \times M}.
\label{eq:23}
\end{equation}

Note that, although the energy harvested from the signals of other sensors $E_{k,\textrm{EH}}$ does not depend on $\bm{Q}(t)$, it is taken into account in the optimization through to the weights $w_k(t)$. Also, it is noteworthy that the scheme designed in this section depends only on the channel statistics, the spatial distribution of sensors and their activity. Hence, since this information is usually known (or can be estimated) in realistic scenarios, a practical implementation of the energy allocation is feasible.

\section{Numerical Simulations}
\label{sec:5}

In this section, we present several simulations to evaluate the performance of the previous approach. To that end, we consider the micro-urban scenario in \cite{ITU09} with $K = 10$, $R_k = 10$ m, $d_{\textrm{min}} = 0.1$ m, $T_f = 1$ s, $T_d = 0.5$ s, $P_{\textrm{tx}} = 40$ dBm, $P_{\tau} = 20$ dBm, $\sigma_{\gamma}^2 = \sigma_g^2 = 1$, $\alpha = 2$ and $p_{\textrm{act}} = 0.1$. The steering vectors  $\bm{v} (\cdot)$ are computed for a uniform circular array \cite{Van04} with $M = 100$. Regarding the density of sensors, in this paper we study three different cases: (i) $\lambda_k = \lambda = 0.1$ $\textrm{m}^{-2}\ \forall k$, (ii)  $\lambda_k$ are equispaced in the interval $[0.5 \lambda, 2 \lambda]$, and (iii) $\lambda_k$ are equispaced in the interval $[0.1 \lambda, \lambda]$.

\begin{figure}[t]
\centerline{\includegraphics[scale=0.43]{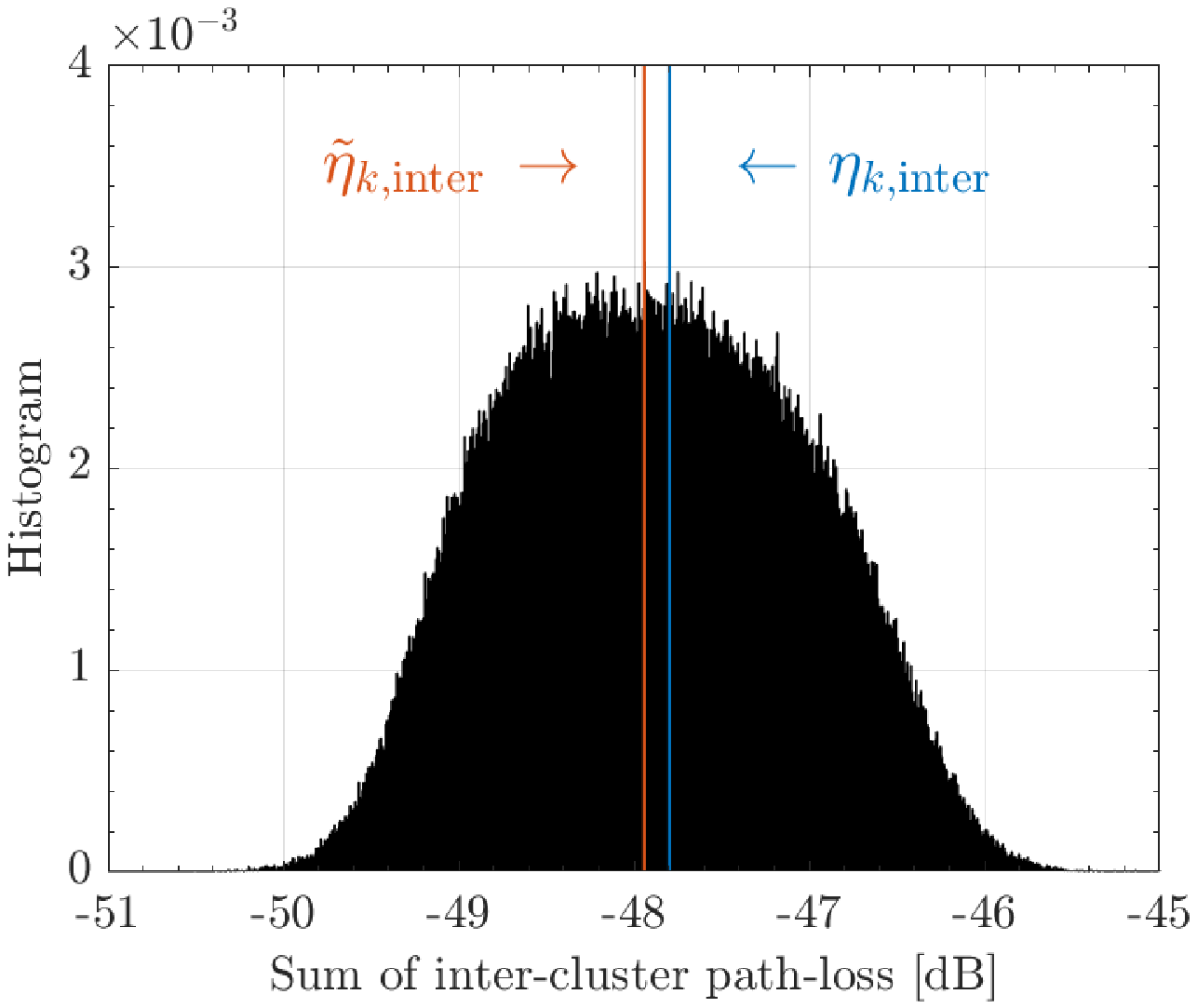}}
\vspace{-0.8em}
\caption{Histogram of sum of inter-cluster path-loss.}
\label{fig:4}
\centerline{\includegraphics[scale=0.43]{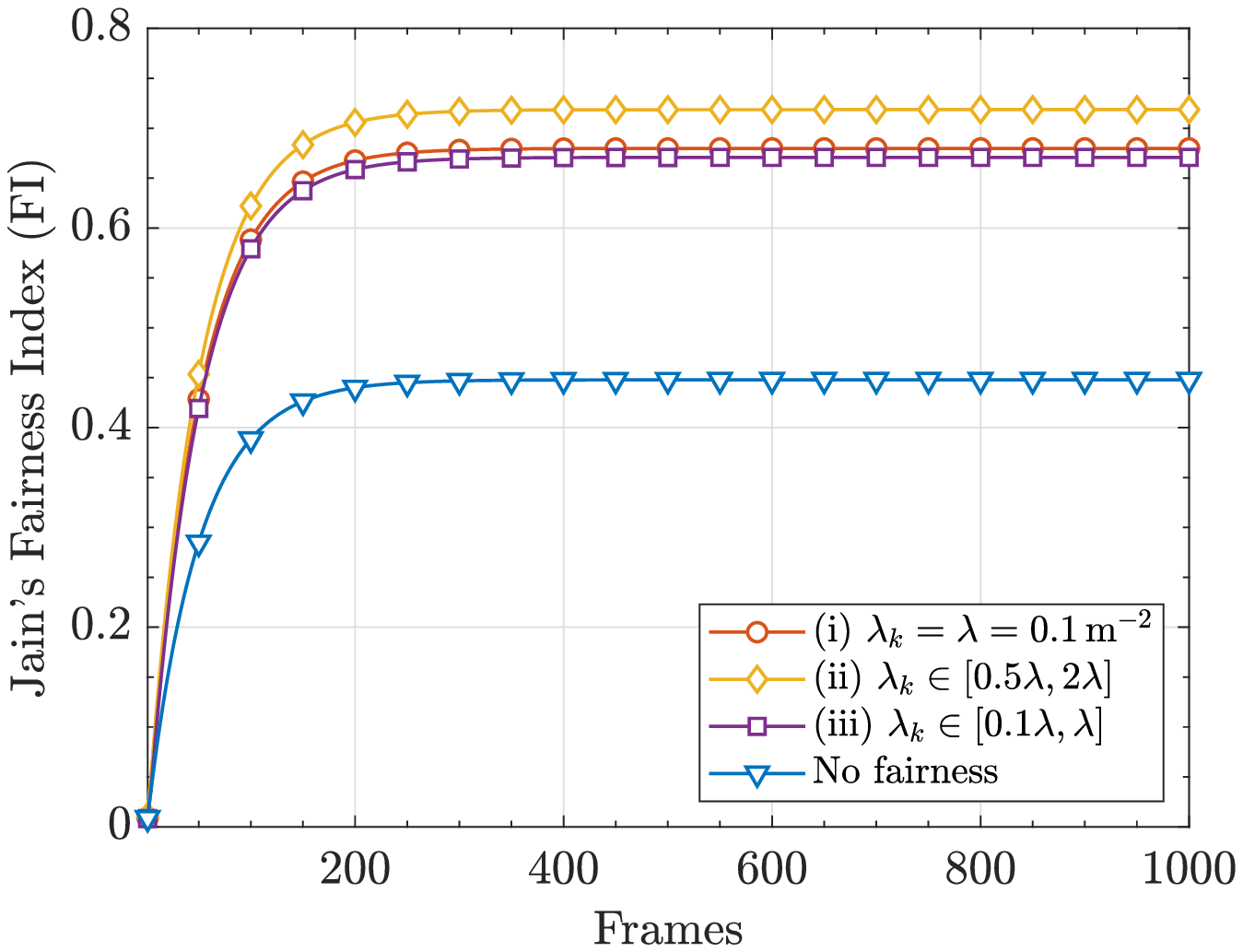}}
\vspace{-0.8em}
\caption{Evolution of Jain's index in time.}
\label{fig:5}
\centerline{\includegraphics[scale=0.43]{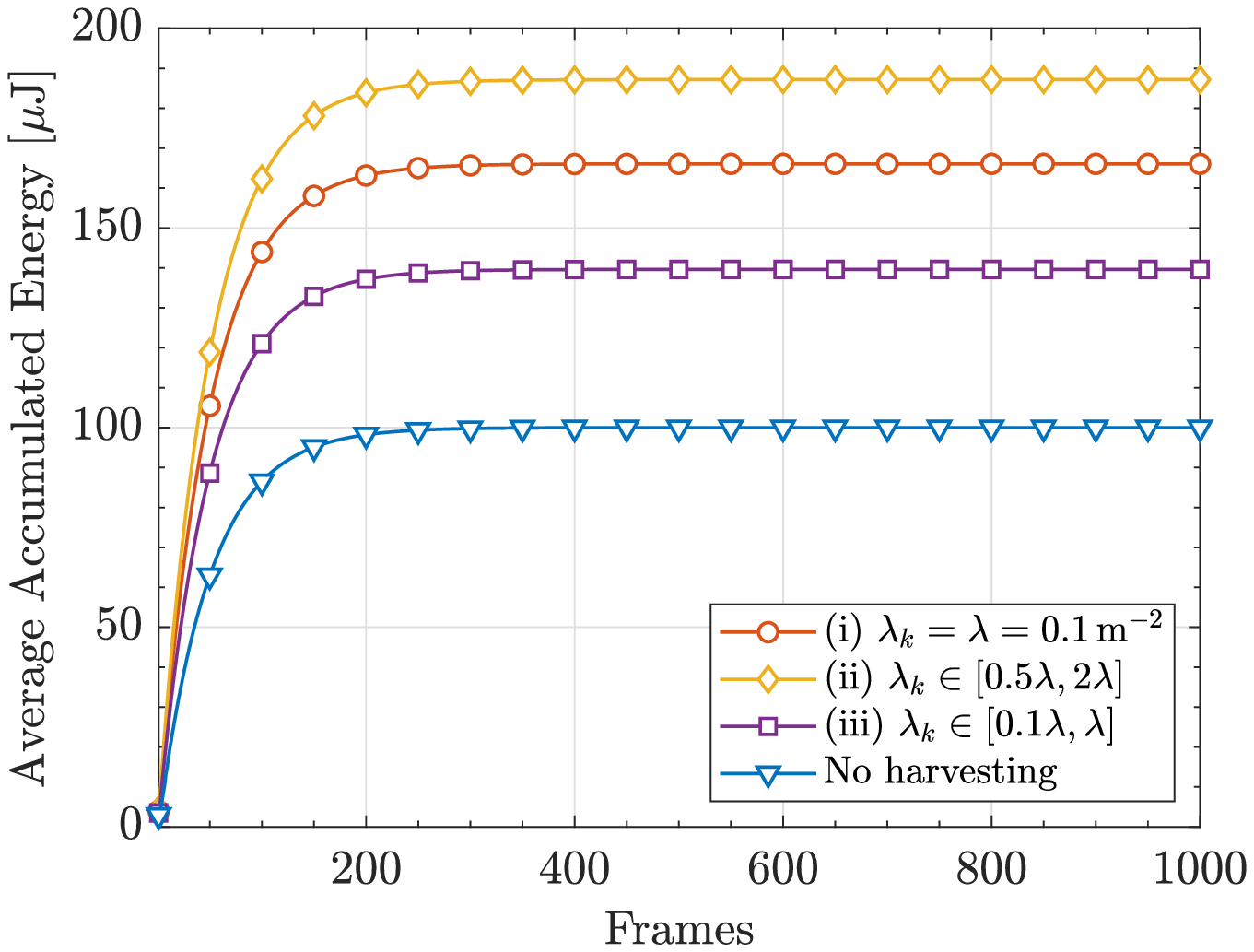}}
\vspace{-0.8em}
\caption{Evolution of average energy in time.}
\label{fig:6}
\end{figure}

As mentioned, we start by validating the approximation of the sum of inter-cluster path-loss in \eqref{eq:17}, i.e. the sum of path-loss at the cluster center is similar over the cluster. For this task, in Fig. \ref{fig:4} we present the histogram of the sum of path-loss together with the approximated value obtained with stochastic geometry $\tilde{\eta}_{k,\textrm{inter}}$ and the actual value $\eta_{k,\textrm{inter}}$. It can be observed that as $\tilde{\eta}_{k,\textrm{inter}} = -47.94$ dB is close to $\eta_{k,\textrm{inter}}= -47.80$ dB, it can be used as a suitable approximation.

To assess the results of the proportional fairness, we consider $T = 1000$ frames with an averaging filter of length $T_c = 50$ frames. In that sense, we make use of Jain's fairness index (FI) to measure the fairness within the system \cite{Tek13,Ngu06}: 
\begin{equation}
\textrm{FI}(t) = \left( 1 - \frac{1}{T_c} \right) \textrm{FI}(t - 1) + \frac{1}{T_c} \frac{(\sum_{k = 1}^K E_k(t))^2}{K \sum_{k = 1}^K E_k(t)^2}.
\label{eq:24}
\end{equation}

Furthermore, regarding the accumulated energy, we also show the average w.r.t. the different frames, defined as %
\begin{equation}
\bar{T}(t) = \frac{1}{K} \sum_{k = 1}^K T_{k} (t).
\label{eq:25}
\end{equation}

Metrics $\textrm{FI}$ and $\bar{T}$ are depicted in Fig. \ref{fig:5} and \ref{fig:6}, respectively. Note that, in order to appreciate the effect of the proportional fairness, in Fig. \ref{fig:5} we also include the case where the sum of stored energies is maximized without a fair policy ($w_k(t) = 1$). Similarly, regarding the impact of the EH, the case with WPT and no EH (no harvesting) is also shown in Fig. \ref{fig:6}.

As illustrated in Fig. \ref{fig:5}, the dispersion diminishes (i.e. FI increases) along time thanks to the proportional fair policy. In fact, after a first transitory period, Jain's index converges to a constant value and, as expected, a poor value is obtained in the case of no fairness in the optimization. 

On the other hand, as shown in Fig. \ref{fig:6}, the case of no harvesting entails a smaller amount of average energy $\bar{T}$. This is not surprising and can be easily seen in the situations with a higher density, namely (ii), where the contribution of the harvested energy is larger. Hence, a considerable improvement is attained when collecting energy from the signals transmitted by other sensors. Besides, note that since the average of (iii) is approximately $0.5 \lambda$, this case results in the lowest energy obtained from the signals of other sensors.

\section{Conclusions}
\label{sec:6}
In this paper, we have addressed the analysis of wireless powered MTC networks when considering EH and WPT as available energy supply mechanisms. In particular, assuming a scenario with a serving multiple-antenna BS and a set of randomly deployed sensors with sporadic activity, we have modeled the collected energy in both EH and WPT. To that end, we have employed repulsive MCPs and Bernoulli RVs. Finally, based on the developed analysis, we have derived an energy allocation scheme under a proportional fair policy to guarantee a fair battery recharging for all sensors over time.

\section*{Appendix}
\label{sec:7}

The proof of the equivalence between \eqref{eq:19} and \eqref{eq:21} follows from the first-order Taylor series approximation, i.e.:
\begin{align}
\, &\underset{\bm{Q}(t) \succeq \bm{0}}{\textrm{argmax}} \, \, \sum_{k = 1}^K \log (T_k(t)) \nonumber \\
= \, &\underset{\bm{Q}(t) \succeq \bm{0}}{\textrm{argmax}} \, \, \sum_{k = 1}^K \log \left(T_k(t - 1) + \frac{1}{T_c}( E_k(t) - T_k(t- 1))\right) \nonumber \\
\stackrel{\textrm{(a)}}{\approx} \, &\underset{\bm{Q}(t) \succeq \bm{0}}{\textrm{argmax}} \, \, \sum_{k = 1}^K \log (T_k(t - 1)) + \frac{1}{T_c} \sum_{k = 1}^K \frac{ E_k(t) - T_k(t - 1)}{T_k(t - 1)} \nonumber \\
= \, &\underset{\bm{Q}(t) \succeq \bm{0}}{\textrm{argmax}} \, \, \sum_{k = 1}^K \frac{E_k(t)}{T_k(t - 1)},
\label{eq:26}
\end{align}
where (a) follows from $\log (x + c) \approx \log(x) + \frac{1}{x} c$. Note that (a) becomes equality for an infinite averaging time $T_c \to \infty$.

\bibliographystyle{IEEEtran}
\bibliography{IEEEabrv,SPAWC2020}

\end{document}